# Operational Design Considerations for Retinal Prostheses

Erich W. Schmid[1,2,3] and Wolfgang Fink[2,3,4]

[1]Institute of Theoretical Physics, Tübingen University, Auf der Morgenstelle 14, 72076 Tübingen, Germany
[2]Visual and Autonomous Exploration Systems Research Laboratory, Division of Physics, Mathematics and Astronomy, California Institute of Technology, Pasadena, CA 91125, USA
[3]Visual and Autonomous Exploration Systems Research Laboratory, Departments of Electrical & Computer Engineering, Biomedical Engineering, Systems & Industrial Engineering, and Ophthalmology & Vision Science, University of Arizona, Tucson, AZ 85721, USA
[4]Doheny Eye Institute, Keck School of Medicine, University of Southern California, Los Angeles, CA 90033, USA

**Abstract**

Three critical improvements for present day and future retinal vision implants are proposed and discussed: (1) A time profile for the stimulation current that leads predominantly to transversal stimulation of nerve cells; (2) auxiliary electric currents for electric field shaping with a time profile chosen such that these currents have small probability to cause stimulation; and (3) a local area scanning procedure that results in high pixel density for image/percept formation (except for losses at the boundary of an electrode array).

## 1. Introduction

Research on retinal implants has been ongoing for about two decades: e.g., [Ch 1993, Hu 1996, Zr 1997, Zr 2002, Ri 2003, Hu 2003, Ch 2004, Li 2004, Pa 2005, Mc 2007, Ho 2008, Hu 2009, Zr 2010]. The idea is to restore a small part of vision to people suffering from blindness due to retinitis pigmentosa or due to age related macula degeneration. A chip with an array of electrodes is placed into an epi-retinal or sub-retinal position. Electric currents emerging from the electrodes are seen by the blind person as small phosphenes. They serve as pixels for presenting an image. Despite some encouraging results the goal of presenting a grayscaled picture with a thousand or more pixels has not yet been reached.

In a preceding paper [Sc 2010] computational tools have been presented, which might be useful in the development of a retinal prosthesis. The applicability of the tools has been demonstrated with several examples. The examples shed some light onto the difficulties of implementing the idea of a retinal prosthesis. In essence these difficulties are:

1. The use of too simple time profiles of the electric stimulation signals. The most common profile is the monophasic square voltage pulse, which yields a biphasic current pulse. Or the biphasic square voltage pulse, which yields a triphasic current pulse. Such pulses lead to the depolarization (or hyperpolarization) of a cell membrane in the first phase of the current, and to a polarization of opposite sign in the next phase of the current. Thus a "yes" is followed by a "no."
2. The idea of using only one electrode per pixel. Several research groups are using a common counter electrode at infinity. One group in Australia is using 6 counter electrodes around a center electrode on a hexagonal grid and, by current splitting, a common counter electrode at infinity as well [Lo 2005]. Counter electrodes at infinity, however, lead to cross-talk [Sc 2010].
3. Too little effort for shaping the electric field (or current). As has been discussed earlier [Sc 2010], simultaneous firing of neighboring electrodes leads to bunching of field lines. This bunching means field shaping, i.e., increased density of field lines above the electrodes. But it also means undesired cross-talk between neighboring electrodes. The Australian group is the first one to perform true field shaping by six so-called "guard electrodes" [Lo 2005].

In [Sc 2010] these difficulties have been illustrated. How to overcome the difficulties is subject of the present paper.

In Section 2 we study the time profile of the electric field (or current). It has been shown in [Sc 2010] that a shorter pulse with a monophasic current, plus a latency period after the pulse, has an advantage over the commonly used pulse types. We will revisit this finding in Section 2.1. In Section 2.2 an entirely different type of stimulation, namely one that excites predominantly by the dielectric (or transient) part of the current will be discussed.

Section 3 will deal with the problem of field shaping. One would like to have an electric stimulation field that becomes active in a small target volume in the retina, without having any major effect elsewhere. Unfortunately, an electric current, and the electric field that drives the current, cannot be focused like a beam of light. What one can do, however, is to use separatrices for guiding the current. This will be discussed in Section 3. The basic element for field shaping will be the dipole ridge, as discussed in Section 3.1. Multipole pixels, with a center electrode and field guiding electrodes around them, will be discussed in Section 3.2. The use of several electrodes for one pixel leads to a reduced pixel density. Local area scanning stimulation will allow us to regain the initial high pixel density, without having to increase the physical number of electrodes, as discussed in Section 3.3. In Section 3.4 a stochastic optimization framework is discussed that can be employed (a) to achieve optimal field shaping with a given electrode array (chip); (b) to optimally design the electrode array itself, i.e., to determine the spatial arrangement and size/diameter of the electrodes on the electrode array; and (c) to optimally drive the stimulation through a given, implanted electrode array in real time during usage of the retinal implant. Concluding remarks are presented in Section 4.

## 2. Optimizing the time profile of a stimulation signal

Research on electrical stimulation of nerve cells started in the late 18[th] century when Luigi Galvani saw frog legs twitch and found out that the twitching was an electrical effect [Ga 1791].

Actually Galvani found two kinds of electrical stimulation that were distinctly different in terms of physics. Since it is important for the present paper, we briefly discuss the two historical experiments.

It started with Luigi Galvani killing frogs and preparing their legs for the kitchen. He did this in his laboratory. At the same time his assistant was working nearby with an electrostatic generator. It happened that Galvani was touching the spine of a frog with his knife when, at the same moment, the electric generator sparked. Amazingly, the legs of the dead frog twitched! Galvani and his assistant repeated the experiment. The frog legs twitched. There was no doubt: the twitching was caused by the spark discharge of the electrostatic generator. In order to give credit to Galvani for his historic discovery we denote this event as a *galvani-1 stimulation*.

Galvani built an antenna and found out that also thunderstorm lightning caused this kind of stimulation. At the time, almost a century before the famous experiment by Heinrich Hertz, Luigi Galvani had no chance of finding out in detail what was going on in this galvani-1 stimulation.

Soon after the first discovery of electrical stimulation there came a second one. This time it happened without any sparking electrostatic generator. Luigi Galvani had hung up cleaned frog legs at an iron grill. When he touched a leg with a piece of copper the leg twitched! Again, Galvani repeated the experiment and modified the setting. The stimulation effect was reproducible. He found out that two different kinds of metal and a closed electric circuit were essential. As we know, this time his research was very fruitful and led to the discovery of the galvanic cell. Again we want to give credit to Galvani for his historic discovery and call the second kind of electrical stimulation a *galvani-2 stimulation*.

These two kinds of stimulation found by Luigi Galvani are distinctly different. What is the basic difference? The electrical process in the galvani-1 experiment is by several orders of magnitude faster than the one in the galvani-2 experiment. The voltage surge may have been stronger in galvani-1 but, due to the extremely short pulse duration (in the nanosecond range!), the transferred electric charge has been by orders of magnitude smaller than in the galvani-2 stimulation.

Today, more than two hundred years later, electrical stimulation is still hard to understand. The reason is that biological tissue is a very complicated electric conductor. It is partly an electrolyte, partly an insulator, partly a colloid. It contains macromolecules with various polarizabilities and it has a complicated structure in space, with cells and clefts in between the cells. Research on its electrical properties has been ongoing ever since Galvani's experiments and is not complete. The history of this research is well documented in a review paper by Foster and Schwan [Fo 1989].

There is another difficulty in a theoretical description of stimulation. The stimulation process of a nerve cell is a non-linear operator in mathematical language. For the analysis of experimental results, Fourier transformations are commonly used [Fo 1989]. A non-linear operator does not commute with these transformations. This means that stimulation cannot be studied in Fourier decomposition! Most experiments on the electric properties of biological tissues, however, are done with alternating currents of various frequencies, i.e., in Fourier decomposition [Fo 1989].

Theoretical descriptions of electric currents in biological tissues and theoretical descriptions of the stimulation process of nerve cells are always based on simplified models, such as continuum models, statistical models, suspension of spherical cells in an electrolyte, etc. The predictive power of such model descriptions is not very high. Nevertheless, the models are useful as a basis for intuitive descriptions, and they are useful for designing experiments. The final answer, of course, will always be given by experiment and not by theory.

In the following we will describe three very different ways of electrical stimulation of nerve cells: (1) the electrical stimulation by a sustained ohmic current, (2) the electrical stimulation by a delta-function-like charge injection, and (3) the stochastic stimulation by a small ohmic current.

## 2.1. Longitudinal stimulation

The standard model of electrical stimulation is based on the findings of Hodgkin and Huxley about the electric properties of nerve cells [Ho 1952] and on Heaviside's cable equation. The original cable equation describes the gradual loss of a telegraph signal in an ocean cable. It also describes the situation when the signal travels in the ocean and the cable picks it up like an antenna; only the driving term of the equation is different in the two cases. When a dendrite or an axon has the mathematical properties of a cable, Heaviside's equation is also valid in a continuum model of a nerve tissue such as the retina.

In [Sc 2010] the antenna-version of the cable equation has been derived, two methods of solution have been presented, and practical examples have been discussed.

Presently, longitudinal stimulation is not in the focus of our interest. We are more interested in transverse stimulation, i.e., perpendicular to the axon or dendrite of a nerve cell. Therefore we only want to shortly review what has been discussed in [Sc 2010] on longitudinal stimulation.

A sustained ohmic current is produced between an electrode and a counter electrode by a so-called voltage-controlled generator or a current-controlled generator. A voltage- controlled generator typically applies a constant voltage of a few volts for a time interval somewhere between 0.3 and 3 milliseconds. The current, after reaching a maximum, goes down while the Helmholtz-layers of the electrodes are charging up. A current-controlled generator keeps the current at a constant value while the applied voltage is adjusted. In both cases, care is taken to avoid irreversible chemical reactions. In the first case, the applied voltage is set accordingly low, in the second case, the injected charge is limited by choosing a small current and a short enough time interval.

Except for the very beginning of charge injection, we will have a sustained ohmic current flowing along the clefts between the cells of the retina. In the underlying continuum model, the clefts and cells of the retina are smeared out, except for the nerve cell under consideration. When the current has a longitudinal (i.e., parallel) component with respect to the axis of the axon or the dendrite under consideration, its electric field will enter into the nerve cell. The process is described by the antenna-version of Heaviside's cable equation. We refer to [Sc 2010] and show a picture of that publication.

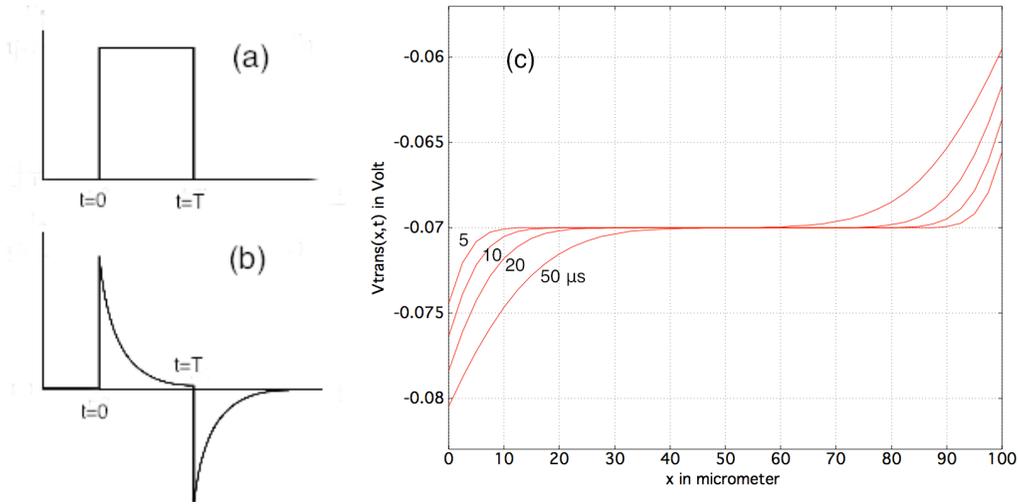

**Fig. 1. Stimulation signal entering into a section of an axon or dendrite of *100 μm* length. (a) the cathodic voltage pulse of *- 2 V* and duration *T=500 μs* applied to the electrode, (b) the time profile of the resulting current (in arbitrary units), (c) the transmembrane potential in the axon or dendrite after *5 μs, 10 μs, 20 μs* and *50 μs*; the resting potential is set to *-0.07 V*.**

In this example, the generator is voltage-controlled and applies a voltage of *-2 V* for a duration of *500* microseconds, as shown in Fig. 1a. Fig. 1b shows the current profile typical for an electrode of small, but sufficient capacity. Fig. 1c shows the solution of the cable equation, for typical cell parameters and a so-called closed end boundary condition. It is seen how the electric field of the ohmic current enters into the cell. A hyperpolarization is produced at one end and depolarization is produced at the other end of the cell section. When the depolarization has reached its critical value the passive cable will become an active one according to Hodgkin and Huxley [Ho 1952], i.e., the stimulation process continues and an action potential arises.

It is interesting to have a closer look at the mathematical model presented in [Sc 2010]. One can see that only the longitudinal (i.e., parallel) component of an electric field, relative to the axis of the axon or dendrite, will have an effect on the solution of the cable equation; a transverse (i.e., perpendicular) component will not have any effect at all.

## 2.2. Transverse stimulation

As has been stated in the preceding section the cable model needs a longitudinal current for the stimulation of a non-myelinated axon or a dendrite. A transverse (i.e., perpendicular) electric current or field cannot stimulate in the cable model. From experiments, however, we know that it can!

In a beautiful experiment, Shelley Fried [Fr 2009] searched for sections of an axon emerging from a retinal ganglion cell that are especially sensitive to electrical stimulation. He placed a conical electrode into the retina, sideways from an axon and pointing toward the axon (i.e., perpendicular to the axon), and injected a constant current for the rather short time of 100-200 microseconds. The transverse current stimulated the axon! How could that happen?

Let us first try to understand what happened in Galvani's first experiment. As has been said, an assistant of Galvani operated an electrostatic generator, charged up the metal spheres until they discharged with a loud spark. At this moment Galvani was still cleaning the frog legs. His iron knife was in contact with the spinal cord of the dead frog. The knife, together with the arm of Galvani, accidentally became an antenna and a very small charge appeared at the tip of the knife. In this way the tip of the knife became a stimulating electrode. In order to understand the electric process leading to this stimulation let us try to visualize things in a slow-motion "movie."

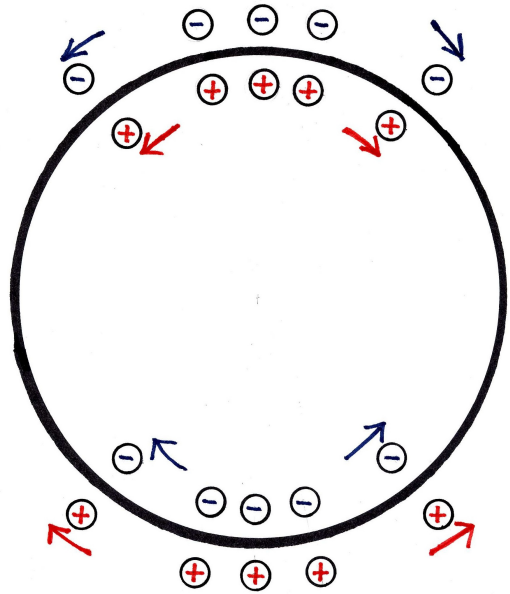

**Fig. 2. Cross-section of a dendrite, or non-myelinated axon, at the time of maximum polarization. Relaxation means that ions are driven sideways, as shown by arrows, and mix with ions from the other side of the circle, thus forming a neutral saline again. While relaxation is going on, the cell membrane is depolarized at one side of the circle and hyperpolarized at the opposite side.**

Let the appearance of a small electric charge at the tip of the knife be the first time frame of our slow motion movie. Of course, the generator cannot deposit the charge in zero time. But it can do so in very short time, i.e., on the order of a microsecond or less.

In a second time frame we consider the Coulomb field of the charge that has appeared at the tip of the knife. This Coulomb field polarizes the retina, or any other biological tissue, that fills the space between the tip of the knife and a remote counter electrode. The ions of the saline, together with their clouds of water molecules, will move a little, driven by the Coulomb field. They will form the Helmholtz layers at the electrode and counter electrode. And they will form double layers at cell membranes, or at any other non-conducting obstacle, as schematically depicted in Fig. 2. Let the stimulating electrode with positive charge be at the bottom of the picture (Fig. 2) and the remote counter electrode at the top and let an axon or dendrite cut the image plane in such a way that the cross section is seen as the circle depicted in Fig. 2. The double layer of charges will be strongest at the bottom and (with opposite sign) at the top of the circle. In the middle between top and bottom there will be no polarization.

Let us look now at further time frames of the slow motion movie. The insulation of the membrane is not perfect, but good enough to keep the ions and counter ions of the double layer apart for some time. The only degree of freedom of these pairs of charges is to move sideways. The driving force for this movement is not very strong, but sufficient to make the charges spread out over the surface of the membrane; at the median of the circle charges of opposite sign mix and form a neutral saline again. The resulting current is called a relaxation current. In the theory of electric currents in biological tissues such relaxation currents play an important role. In fact, there exist several relaxation currents with different logarithmic decay times [Fo 1989]. Here we are only interested in the relaxation process that plays the dominant role in transverse stimulation of nerve cells.

Let us recall the frames of our slow motion movie: *charge injection – polarization of membranes – relaxation current.* It is a simplified model. We use it in order to illustrate hyperpolarization and depolarization of the membrane of a nerve cell by a transverse electric field. In reality, the time frames overlap, and a small ohmic current will accompany the displacement current.

Nevertheless, the simplified model of our slow motion movie is useful. It gives us an explanation of what Luigi Galvani has likely seen in his very first experiment. As has been said above, the hand of Galvani and the knife in his hand acted like an antenna and, in less than a microsecond, a very small electric charge appeared at the tip of the knife. A displacement current ran through the tissue of the frog and polarized the membrane of ganglion cells. There was no counter pulse, like in an AC current. The polarization decayed according to relaxation. Apparently, depolarization of parts of the cell membrane during relaxation, followed by a latency period, was sufficient to trigger an action potential that let the frog legs twitch.

We come back now to the experiment by Shelley Fried and his collaborators [Fr 2009]. The cable model does not explain their findings. There has been an ohmic current for 100-200 microseconds. But this current was aiming at the spot in which the investigators were interested, i.e., had a direction perpendicular to the axon. Farther away from that spot it did have a longitudinal component, but the chosen polarity yielded hyperpolarization instead of depolarization. This precludes longitudinal stimulation. The sharp onset of charge injection with the rectangular time profile of the current, however, produced a sufficiently strong voltage surge for a galvani-1 or transverse stimulation. The rest period needed for the galvani-1 stimulation had been set to 10 milliseconds and was part of the time profile.

Thus, for the operation of a retinal implant, transverse stimulation seems to be an excellent alternative to longitudinal stimulation. An injected charge-controlled generator, instead of a voltage-controlled or current-controlled one, would be ideal. It is important that the charge is injected very fast – termed flash or shock stimulation – faster than relaxation, and without a voltage bound. This means that the energy needed has to be stored before injection in the circuitry of the generator (e.g., in a capacitor). Injection may be repeated after the relaxation current has decreased from its maximum value to some given lower value, before discharging the electrode. Electric field shaping may be needed to hit the target cells (see below).

## 2.3. Stochastic stimulation

Stochastic stimulation of the retina takes place when the stimulation current in extracellular space interferes with the omnipresent synaptic noise of a neural network. Synaptic noise is known from the cortex of the brain; it is expected to be present also in the ganglion cell layer of the retina. The word noise means "non-directional" or "no transport of information". A small electric field may be sufficient to make such noise "directional", i.e., transporting information. In the retina such primitive information might be perceived as a phosphene. There is a similarity to the Maxwell-Boltzmann velocity distribution of gas molecules: a very small deviation from a spherical, isotropic distribution is macroscopically observed as "wind", i.e., directional. The magnitude of an electric current sufficient to cause a phosphene is expected to be small, smaller than the current needed for longitudinal stimulation. We expect such phosphenes to appear, for instance, in area C of Fig. 3.

## 2.4. Mixed longitudinal and transverse stimulation

In most experimental applications one has a mixture of longitudinal and transverse stimulation. The time profile shown in Fig. 1a, for instance, starts with a charge injection at the beginning and goes over into an exponentially decaying ohmic current, before discharging starts at t=T. If we take it as a superposition of a signal for longitudinal stimulation plus a signal for transverse stimulation, the ohmic current phase can be viewed as the latency period of the signal for transverse stimulation.

If we want to measure pure (or almost pure) longitudinal stimulation we have to start with a voltage ramp, instead of a voltage step, to avoid transverse stimulation. Even in that case, some transverse stimulation might occur while we are going up the ramp because of a non-zero time derivative of the voltage. This is why pure longitudinal stimulation is hard to investigate.

If we increase the value of T in Fig. 1a and get as a result an increased stimulation probability, for instance, it is hard to tell whether this increase is due to longer latency of galvani-1 stimulation, or due to the longer ohmic current phase in a galvani-2 stimulation.

It is easier to design an experiment for pure galvani-1 or transverse stimulation. We only have to apply a charge injection that is shorter than the time constant of the relaxation current. Modern electronics can do that without Galvani's electrostatic generator. It might be helpful to measure all properties of pure transverse stimulation first, before trying to analyze the stimulation effect of the mixed signal.

## 3. Optimizing the spatial profile of a stimulation signal

When we talk about electrical stimulation of the retina we tacitly assume that stimulation takes place in a prescribed target volume. In [Sc 2010] examples of such target volumes are shown for the case of one activated electrode in a sub-retinal position, a counter electrode at infinity, and the vitreous replaced by silicon oil. Fig. 1.2 in [Sc 2010] is presented here as Fig. 3.

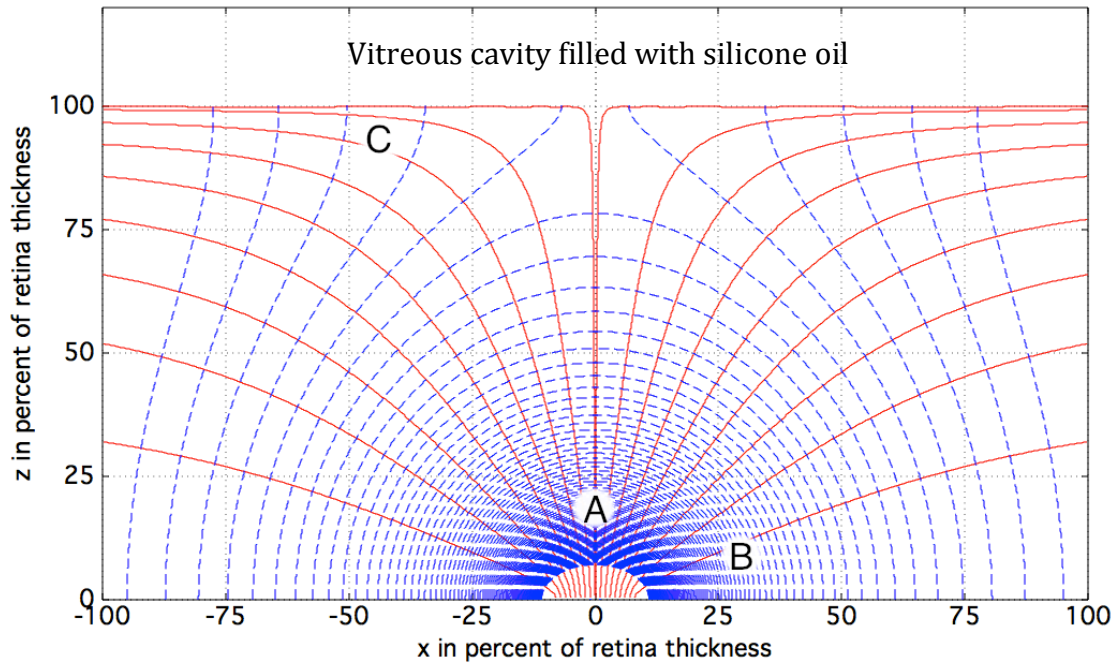

Fig. 3. A typical stimulation field produced by one activated electrode in sub-retinal position. The counter electrode is at infinity, the vitreous is replaced by silicon oil. The electric current field is shown in solid red lines, the equipotential electric field is shown in broken blue lines; the density of blue lines is proportional to the strength of the current. The letters A, B and C mark typical target volumes for the stimulation of bipolar cells, dendritic connections, or neural networks in the ganglion cell layer, respectively.

The area around A is the ideal target volume for the stimulation of bipolar cells by a longitudinal field. In the target volume B one finds bipolar cells for transverse stimulation or dendritic connections. In the target volume C one expects to find neural networks of the ganglion cell layer. Depending on the type of stimulation one would like to "aim" at such target volumes, i.e., one would like to have a strong stimulation field in that volume and a less strong field elsewhere.

Unfortunately, an electric current cannot be focused like a laser beam. The electric current field has to satisfy the Poisson equation, and there is a non-crossing rule of field lines; what is true for electric currents is also true for electric fields. If we want to do some field shaping we have to look for means that are consistent with these rules.

## 3.1 Field shaping using separatrices

When an electric field or a current field is produced by an array of electrodes, every field line goes from exactly one of the anodes to one of the cathodes. A neighboring line at infinitesimal distance goes from the same anode to the same cathode. There are areas, however, in which field lines emerge from another anode and/or go to another cathode. Every two of such areas are separated from each other by a mathematical surface called "separatrix" [Le 1990]. The position and shape of such separatrices are determined by the geometry of electrode positions and by the activation potentials. In the following we want to use such separatrices for shaping the electric current of our stimulation signals.

### 3.1.1 Separatrix formed by an arrangement of dipoles

As an example we construct a separatrix that has the form of a ridge. It is formed by an arrangement of electric dipoles, see Fig. 4a. The direction of the dipoles is perpendicular to the ridge. The ridge has peaks and saddles. The stimulation electrode is situated in front of the ridge, its counter electrode is situated behind the ridge. The surface of the ridge is the separatrix for the stimulation current. The field between stimulation electrode and counter electrode cannot penetrate into the ridge: it has to climb over it; the current density will be high near a saddle. An example of field lines is shown in Fig. 4b. The highest point of the separatrix is a saddle in this figure. The field lines in close vicinity of the stimulating electrode are not shown for plotting reasons. The field lines between the electrodes of the dipoles (shown in blue in Fig. 4a) fill the space between the separatrix and the x-axis, i.e., underneath the separatrix. They are considered being an auxiliary field and are not shown in Fig. 4b.

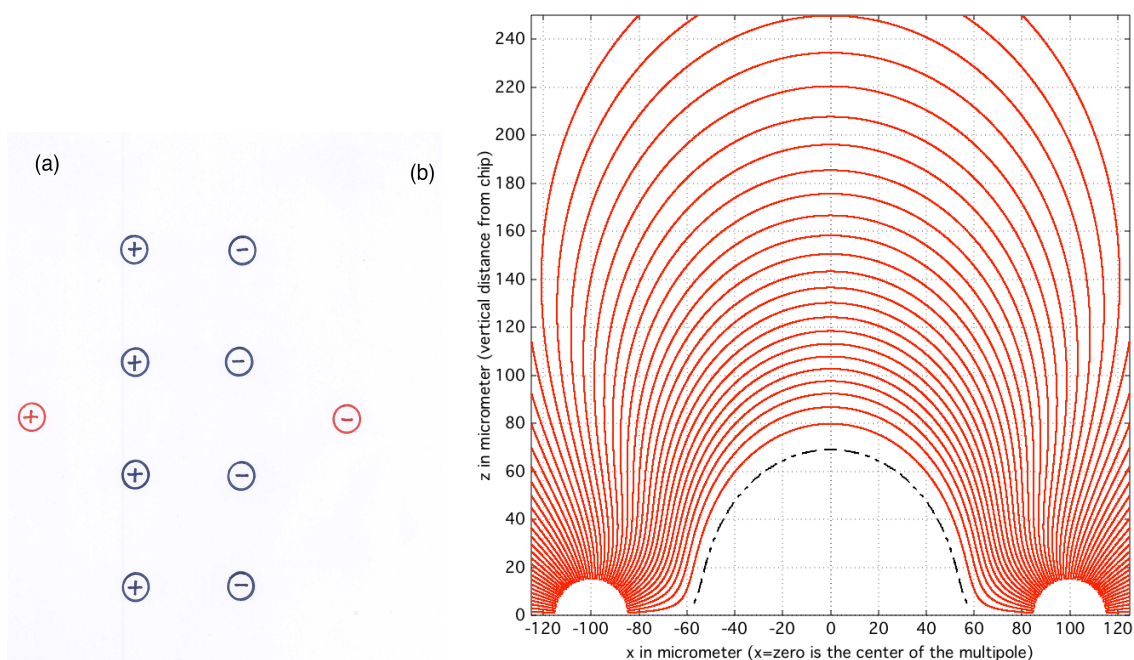

**Fig. 4** A separatrix that has the shape of a ridge for the stimulation current is built by 4 dipoles. In (a) a top view of the relevant electrodes on the array is seen. The blue electrodes are the 4 dipoles producing the ridge, the red electrodes are the stimulation electrodes. In (b) the stimulation field is seen in red on a plot screen vertical to the array with a base line including the stimulation electrodes. The broken line is the separatrix. The semicircles around the stimulating electrodes have no physical meaning; they are introduced for plotting reasons. The field lines of the electrodes shown in blue are not plotted.

The auxiliary field is not supposed to interfere with the stimulation. This is achieved by giving it another time profile. For predominantly galvani-1 stimulation, for instance, this can be done by choosing a time profile of the auxiliary current that yields a predominantly galvani-2 stimulation. For didactical simplicity let us take a narrow gaussian shape for the stimulation field and a wider gaussian shape for the auxiliary field. The width of the stimulating gaussian might be the logarithmic decrement of relaxation; compare to the discussion of Fig. 2. The width of the gaussian profile of the auxiliary field might be three times larger. The charge injection of the galvani-1 stimulation is very small, and, although the charge injection of the galvani-2 stimulation is several times larger, the auxiliary current is still too small to cause heat damage of the tissue. The ridge becomes higher when the electrodes forming the ridge emit more current. It becomes lower when they emit less current. In Fig. 4b the situation is shown for the case that all electrodes emit/receive the same amount of current.

### 3.1.2 Multipoles formed by stimulating electrodes and field shaping electrodes

An example of using several electrodes of an array for guiding a stimulating current to a prescribed target volume is seen already in Fig. 4. We need at least 2 dipoles, i.e., 4 electrodes, to form a guiding field plus 2 electrodes for stimulation. The configuration is suitable for stimulation in target area B of Fig. 3. The distance of the target volume from the chip can be increased either by increasing the strength of the guiding field or by using electrodes that are farther apart from each other.

In order to stimulate in target area A of Fig. 3, with a vertical current, we need a center electrode encircled by ridge forming dipoles. We see such a configuration in Fig. 5: The chip carries a hexagonal grid of electrodes as used in experiments by the Australian group [Lo 2005]. Again, the electrodes forming the separatrix for the stimulation current are shown in blue. The stimulating electrode in the center and the return electrodes are shown in red.

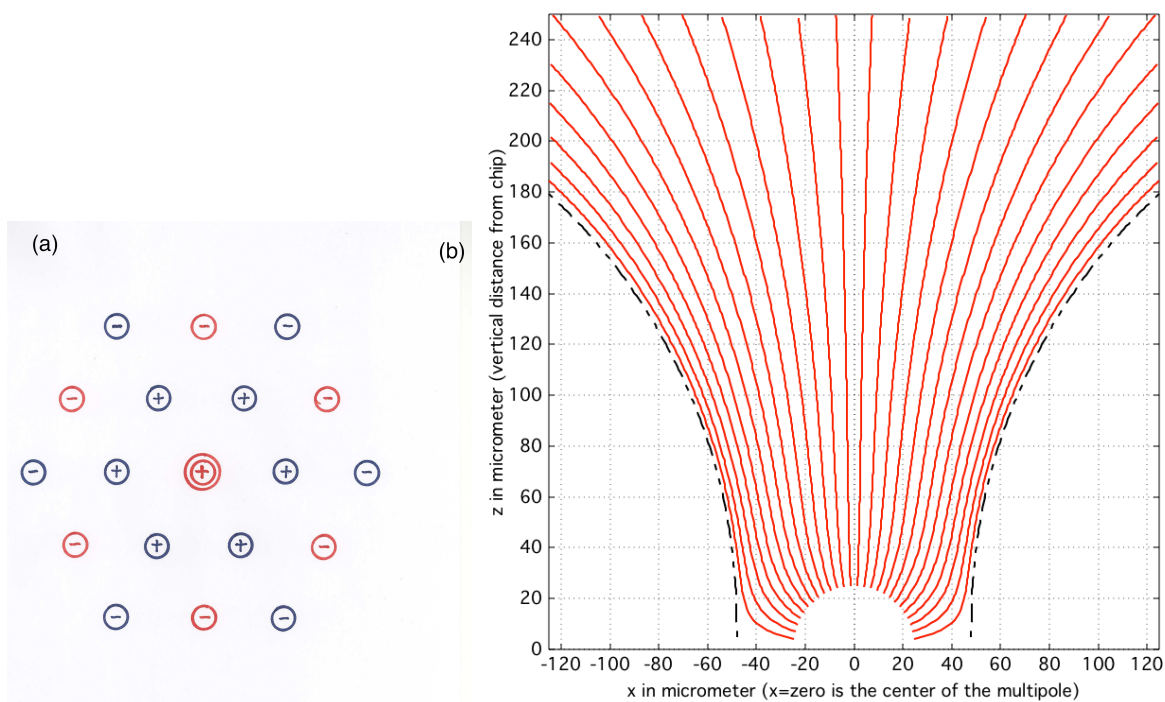

**Fig. 5. A multipole consisting of 19 electrodes, in a hexagonal array. The multipole represents one pixel. The stimulating current emerges from the center electrode and goes to the six counter electrodes shown in red. The electrodes shown in blue build up a separatrix for guiding the stimulation current. The signs shown are given for anodic stimulation; for cathodic stimulation the signs have to be reversed.**

The stimulation current can be viewed as a fountain, rising in the middle, dividing up and falling down to the return electrodes; in the numerical simulation (Fig. 5b), the electrode has been substituted by an adequate boundary condition on a hemisphere.

What has been done for a hexagonal grid of electrodes can also be done for a cubic one as shown in Fig. 6. Again, dipoles around a center electrode are forming a ridge for the stimulating current.

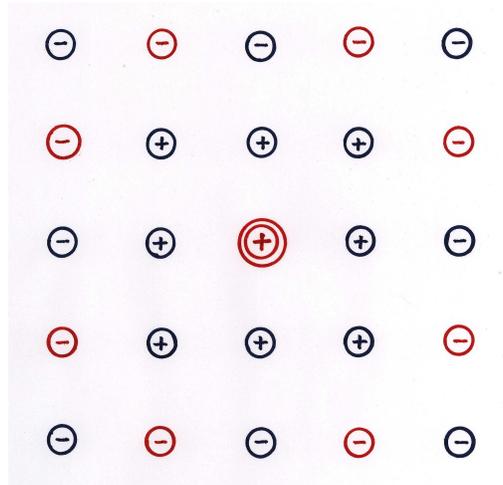

Fig. 6. A multipole consisting of 19 electrodes, in a quadratic array. The multipole represents one pixel. The stimulating current emerges from the center electrode and goes to the six counter electrodes shown in red. The electrodes shown in blue build up a separatrix for guiding the stimulation current. The signs shown are given for anodic stimulation, for cathodic stimulation the signs have to be reversed.

## 3.2 Local area scanning stimulation

The idea of imaging with an array of electrodes is to utilize as many pixels as there are electrodes, and not to use many electrodes for shaping an electric current field. The latter is done in Fig. 4, and more so in Figs. 5 and 6. In the present subsection it will be shown how to get back to the basic idea.

Research on retinal implants is carried by the hope that the phosphenes produced by electrical stimulation are small and thus can be used as picture elements called "pixels". It would be ideal to have as many pixels as there are electrodes on the chip.

In Fig. 5a and in Fig. 6 we are using 19 and 25 electrodes, respectively, for producing only one pixel. The center electrode is used for transmitting the stimulation signal and 24 electrodes are used partly as counter electrodes and partly for shaping the stimulation field. At first sight we seem to be using too many electrodes for producing one single pixel.

During the operation of a retinal implant we do not want to transmit only one single image but obviously a video sequence of images. We have to look at the time scale. Typically, a video sequence has 25 frames per second. This means that there are 40 milliseconds time for transmitting one frame. This is more time than needed for completing an electrical stimulation process.

For longitudinal stimulation the typical length of a monophasic voltage pulse is 0.5 milliseconds, as in Fig. 1a. The resulting biphasic current pulse decays to a current density below stimulation threshold after about 0.8 milliseconds, see Fig. 1b. The membrane needs some time to respond. If we give the same amount of time for that, the stimulation process should be completed after about 1.6 milliseconds. For transverse stimulation the stimulation process needs about a *microsecond* for charge injection and a rest period of up to 1 millisecond. Thus we need a time window of 1.6 milliseconds for completing a process of longitudinal stimulation and a little less for transverse stimulation. This means that we have at least 25 time slices for every one of the 40 millisecond frames of the video sequence. After each one of these 25 time slices the allocation of electrodes to multipoles can be changed. Except for the boundary of the array, every electrode can become the center of a multipole like the ones shown in Figs. 5a and 6. The change of allocation is illustrated in Fig. 7, as an example. In this way we recover the full resolution of the electrode array, except for a certain loss at the boundary of the array. This loss along the circumference of an electrode array would favor circular over rectangular electrode arrangements, which would also conform better to the fundus. In case of the multipole shown in Fig. 6 the jumping sequence may be row by row, akin to a television screen.

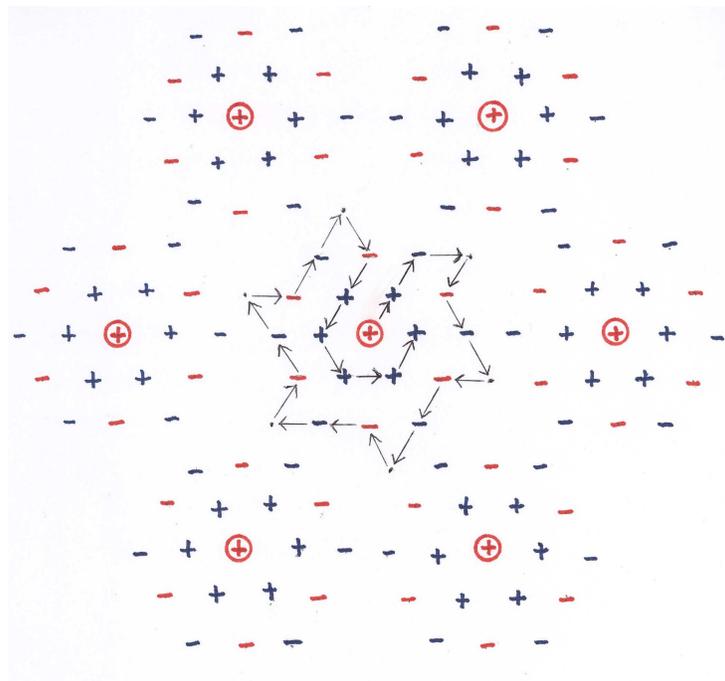

**Fig. 7. Example of local area scanning with multipoles of the type shown in Fig. 5a. The centers of the multipoles jump after each time slice to a new position on the hexagonal grid, as shown by arrows for the multipole in the center. The sequence of jumping is irrelevant as long as it covers all electrodes within the time between two consecutive video frames.**

## 3.3 Transverse, monopolar stimulation with a remote counter electrode

So far we did not consider the case in which the desired stimulation volume is situated right on top of the electrode, has a diameter not larger than the electrode and a thickness of less than 100 μm. In this case, unipolar stimulation with one large remote return electrode seems to be feasible, provided that galvani-1 stimulation according to Section 2.2 is employed.

Again the criteria for a galvani-1 stimulation are as stated above: (1) an extremely short charge injection (in the range of nanoseconds, up to a few microseconds), (2) a very small amount of injected charge (of less than 1 nC) and (3) a rather long rest period (of about 1 millisecond) after each charge injection. The occurrence of cross-talk during the short time of charge injection can easily be avoided. Some cross-talk may arise when a second electrode in close neighborhood fires while the first electrode is still in its rest period. This can be avoided by randomizing the firing order, or, in computer-controlled applications, it can be avoided by a computer algorithm that governs the stimulation sequence.

## 3.4 Spatio-temporal optimization of electrode array design and electric field shaping via stochastic optimization framework

In this Section 3.4 a stochastic optimization framework is discussed that can be employed for the spatio-temporal optimization of the electrode array design (i.e., the spatial arrangements of the electrodes on the chip) and the process of electric field shaping. The goals are: (a) to achieve optimal field shaping with a given electrode array (i.e., chip), (b) to optimally design the electrode array itself, i.e., to determine the spatial arrangement and size/diameter of the electrodes on the electrode array, and (c) to optimally drive the stimulation through a given, implanted electrode array in real time during usage of the retinal implant.

In 2008 one of us (WF) has introduced a Stochastic Optimization Framework (SOF) [Fi 2008a] that allows optimizing a system or process that is governed by numerous adjustable parameters. The underlying principle is the minimization of a fitness function that measures the difference between a desired outcome and an actual outcome while operating the system or performing the process. The minimization of this fitness function is accomplished via multi-dimensional stochastic optimization algorithms, such as Simulated Annealing [Me 1953; Ki 1983], Genetic Algorithms [Go 1989], and other Evolutionary Algorithms. The common characteristic of these algorithms is the capability to escape local minima of a fitness function while approximating (and ideally reaching) the global minimum. This characteristic is in stark contrast to deterministic, gradient-descent-based algorithms, such as the Marquardt-Levenberg algorithm [Pr 1991], which tend to be poor performers in multi-dimensional landscapes that exhibit multiple local minima and are locally rugged. The stochastic optimization algorithms use as their input the fitness function value for each iteration and generate a new, modified set of parameter values as a result, ultimately converging to a set of parameter values that yields the desired fitness function value, i.e., close to zero.

The Stochastic Optimization Framework has been applied to the optimization of prosthetic vision, in particular provided by epi-retinal implants [Fi 2008b, Fi 2010a], to manipulate electric stimulation parameters of an implanted epi-retinal electrode array to optimize the resulting visual perception of the blind subject.

In a similar manner, the SOF can be applied to the following optimization scenarios:

1. ***Optimal field shaping with a given electrode array (i.e., chip):***
   A given electrode array means that the spatial location and specifications for each electrode (e.g., diameter, electrode material) are fixed. The SOF can now be employed to optimize the current through the various electrodes such that the resulting 3D shape of the electric field approximates within user-defined tolerances the desired one. The 3D shape of the electric field can be accurately simulated as described in [Sc 2010]. The underlying electrostatic model of the electrode array will deliver the fitness function necessary for the SOF.
2. ***Optimal electrode array design:***
   This scenario is an extension of the first one above: Here the constraint of having a spatially fixed electrode arrangement is relaxed, and the SOF is employed twice: (1) in an outer optimization loop, optimizing the spatial location/arrangement and even the specifications for each electrode (e.g., electrode diameter), and (2) in an inner optimization loop, performing exactly the electric current optimization as described in scenario 1. The result will be an optimized electrode array design.
3. ***Real-time stimulation optimization:***
   Using a sufficiently capable miniaturized computing system (e.g., [Fi 2010b]) one can perform the SOF-based optimization described in scenario 1 in real time (i.e., during actual usage of the vision prosthesis) given the fixed geometry and specifications of the implanted electrode array and the electrostatic model of the electrode array to generate the fitness function for the SOF. With prior image processing of the camera images that feed into the artificial vision implants [Fi 2010b] (especially retinal implants) one can estimate what the desired 3D shape of the electric field for each individual processed camera frame should be. This makes a comparison of the resulting 3D shape of the electric field across the electrode array during the SOF-based optimization process and the desired 3D shape possible, thus delivering the fitness function. Note, that all of this happens prior to the actual electric stimulation of each camera frame via the electrode array: only the respective current profiles of all electrodes underlying the sufficiently converged 3D shape of the resulting electric field corresponding to the image-processed camera frame will be stimulated before the next camera frame is processed accordingly.

As the details of these optimization scenarios are beyond the scope of this work, they will be subject to a forthcoming paper including optimization examples.

## 4. Summary and discussion

Three different mechanisms of electrical stimulation of the retina have been discussed.

The *first* one is called longitudinal stimulation or galvani-2 stimulation. It arises when an ohmic current produces a voltage drop along its course through the tissue of the retina. The component of the voltage drop parallel to a non-myelinated axon or dendrite enters through the cell membrane into the cell as described by the cable equation. This process is well known and used in many applications. It corresponds to the second frog leg experiment by Luigi Galvani.

The *second* mechanism is called transverse stimulation or galvani-1 stimulation. It arises when a very small electrical charge appears suddenly at the surface of an electrode and sends a displacement current through the tissue. Polarizing charges appear at the surface of every insulator, and also at the surface of cell membranes. Since the membranes are surrounded by an electrolyte the surface charges disappear by relaxation. During the time of relaxation the membrane of a nerve cell can react by opening ion channels causing action potentials. We call this process transverse stimulation because the field component perpendicular to the cell membrane is causing the polarization; because of the shortness of the electric pulse we refer to this process also as flash or shock stimulation. It corresponds to the first frog leg experiment by Luigi Galvani.

The *third* mechanism is stochastic stimulation by a very small electric current. Synaptic noise in a neural network does not carry information. Signals travel statistically between neurons without directional preference. A small electric current can impose a directional preference and lead to a net signal. It may cause the perception of phosphenes. One does not expect such phosphenes to be small in size and suitable for imaging. Therefore one should try to avoid this mechanism.

It would be ideal to "focus" the stimulation current to a small target volume and to stimulate neurons of the retina only in this volume. Unfortunately, an electric current has a "focus" only at sources or sinks. However, so-called separatrices can be used in order to conduct some field shaping. It has been discussed how a separatrix can be used to create something like a mountain pass: An auxiliary field creates a ridge with summits and passes. The stimulation current cannot penetrate into the auxiliary field, because of the separatrix between the two fields. It has to climb over the ridge, preferably over a pass. At a pass the field lines are more densely packed. This may be considered a substitute for "focusing". The drawback is that one needs several electrodes of the implanted electrode array (chip), in addition to the stimulating electrode, for this kind of field shaping. At first sight, one sacrifices many electrodes for creating imaging pixels, which means loosing image resolution.

It has been discussed how image acuity can be recovered via local area scanning. One knows how much time is needed for communicating one frame of the video sequence, and one knows how much time is needed for stimulating of a nerve cell. It turns out that the time needed for one video frame can be divided up into time slices. In every one of these time slices the multipole can change position on the chip. In one case there are 25 electrodes needed for creating a stimulating current plus guiding currents. With 25 time slices the center of the 25-pole can move over 25 positions and produce 25 pixels, thus recovering the full image resolution; some loss arises only at the circumference of the electrode array.

The operational design of a retinal vision implant becomes simple when the target volumes are located directly above each electrode and when galvani-1 stimulation is employed. The only difficulty in this case is a technical one: One has to randomize the time sequence of stimulation and the circuitry has to be capable of producing the very short voltage surge needed for the galvani-1 stimulation.

It should be emphasized that the described stimulation modalities (i.e., galvani-1 and glavani-2 stimulations) and the electric field shaping technique via separatices are potentially applicable to the stimulation of other tissue/nerve regions, e.g., optic nerve, lateral geniculate nucleus (LGN), visual cortex, deep brain, and paralyzed limbs.

**Acknowledgement:** Fruitful discussions with Robert Wilke are gratefully acknowledged.

**Disclosures:** Authors EWS and WF may have proprietary interest in the stimulation modalities presented in this manuscript as provisional patents have been issued.